\documentclass[a4paper,12pt]{article}
\newtheorem{proposition}{Proposition}

\usepackage{amsfonts}
\usepackage[round]{natbib}

\usepackage{amssymb,amsmath}
\usepackage[dvips]{graphicx}

\usepackage[margin=2.5cm]{geometry}

\begin{document}
\date{}
\author{D. Salmer\'on$^{1, 2}$ and J.A. Cano$^3$\\
\textit{\small{$^1$CIBER Epidemiolog\'ia y Salud P\'ublica-CIBERESP.}}\\
\textit{\small{$^2$Departamento de Ciencias Sociosanitarias,
Universidad
de Murcia, Spain.}}\\
\textit{\small{$^3$Departamento de Estad\'istica e Investigaci\'on
Operativa, Univeridad de Murcia, Spain.}}\\
}


\title{\textbf{Monte Carlo error in the Bayesian estimation of risk ratios using log-binomial regression models: an efficient MCMC method}}

\maketitle

\begin{abstract}
In cohort studies binary outcomes are very  often analyzed by logistic regression. However, it is well-known that when the goal is to estimate a risk ratio, the logistic regression is inappropriate if the outcome is common. In these cases, a log-binomial regression model is preferable. On the other hand, the estimation of the regression coefficients of the log-binomial model is difficult due to the constraints that must be imposed on these coefficients. Bayesian methods allow a straightforward approach for log-binomial regression models, produce smaller mean squared errors and the posterior inferences can be obtained using the software WinBUGS. However, the Markov chain Monte Carlo (MCMC) methods implemented in WinBUGS can lead to a high Monte Carlo error. To avoid this drawback we propose an MCMC algorithm that uses a reparameterization based on a Poisson approximation and has been designed to efficiently explore the constrained parameter space.
\newline


\end{abstract}

\section{Introduction}

The odds ratio is a measure of association widely used in Epidemiology that can be estimated using logistic regression. On the other hand, when one wants to communicate a risk ratio, the logistic regression is not recommended if the outcome is common, see \cite{mcnutt}, \cite{pedersen2003}, \cite{greenland2004}, \cite{Spiegelman2005}, \cite{pedersen}, and \cite{deddens} among others.

If one wants to estimate the adjusted risk ratio, a log-binomial model is preferable to a logistic model. The log-binomial model assumes that the distribution of the outcome $y_i$ is the Bernoulli distribution 
\begin{equation}
y_i\sim Ber(p_i),\,\,\,
\log p_i=x_i\beta,\,\,\,i\in\mathbb{N}_n=\{1,...,n\},\label{Modelo}
\end{equation}
where $x_i\beta=(x_{i1},x_{i2},...,x_{ik})(\beta_1,...,\beta_k)^T$, and $x_i$ includes variables denoting exposures, confounders, predictors and product terms. Usually $x_{i1}=1$ and therefore $\beta_1$ is the intercept. Since $p_i=\exp(x_i\beta)\in(0,1)$, we have to impose the constraints $x_i\beta<0$, $i\in\mathbb{N}_n$, on the values of $\beta$ which complicates its maximum likelihood estimation. \cite{zou2004} and \cite{Spiegelman2005} have suggested a Poisson model without the constraints, that is,
\begin{equation}
y_i\sim Poisson(\mu_i),\,\,\,\log \mu_i=x_i\beta,\,\,\,i\in\mathbb{N}_n,\label{Zou_poisson}
\end{equation}
to approximate the log-binomial maximum likelihood estimator and they consider a robust sandwich variance estimator to estimate the standard errors. Model (\ref{Zou_poisson}) can be fitted with standard statistical packages like R, STATA or SAS. Nevertheless, if $\hat{\beta}$ is the estimate obtained fitting the Poisson model then $x_i\hat{\beta}$ can be greater than zero. On the other hand, \cite{pedersen}, and \cite{deddens} have proposed a different approximation using an \textit{expanded dataset} and a maximun likelihood estimator.

In this article we consider a Bayesian analysis of the log-binomial regression model (\ref{Modelo}). In this context \cite{chucole} have proposed to incorporate the constraints $x_i\beta<0$, $i\in\mathbb{N}_n$ as part of the likelihood function for log-binomial regression models and they have shown that the Bayesian approach provides estimates similar to the maximum likelihood estimates and produces smaller  mean squared errors. Posterior computations can be carried out using the WinBUGS code that appears in \cite{chucole}; however, WinBUGS can lead to a poor convergence and a high Monte Carlo error. Furthermore, the instrumental distribution implemented in WinBUGS to simulate each full conditional distribution has not account for the constraints in an efficient way: the constraints must be evaluated \textit{a posteriori} each time a simulation from the instrumental distribution is proposed and this simulation is rejected if the new proposed value of the parameter does not satisface the $n$ constraints.

In this paper we overcome these two drawbacks using an MCMC method based on a reparameterization and an instrumental distribution that directly generates values of the parameters in the constrained parameter space.

\section{Simulation from the posterior distribution}

To introduce the problems that can arise we consider the following example discussed in \cite{chucole}. The data are $\mathbf{y}=(0, 0, 0, 0, 1, 0, 1, 1, 1, 1)$ and $x_{i2}=i$, so that
\begin{equation}
\log p_i=\beta_1+i\beta_2,\,\,\,i\in\{1,2,...,10\}.\label{toy}
\end{equation}
We have used the WinBUGS code proposed by \cite{chucole} and provided in the Appendix with the prior distribution $\pi(\beta_1,\beta_2)=1$. We have run a Markov chain with 10000 iterations and an adaptive phase of 500 iterations using the method \textit{UpdaterMetnormal}, and the last 9500 iterations have been used to carry out the inferences.

\begin{figure}[h]
\centering
\includegraphics[scale=0.45]{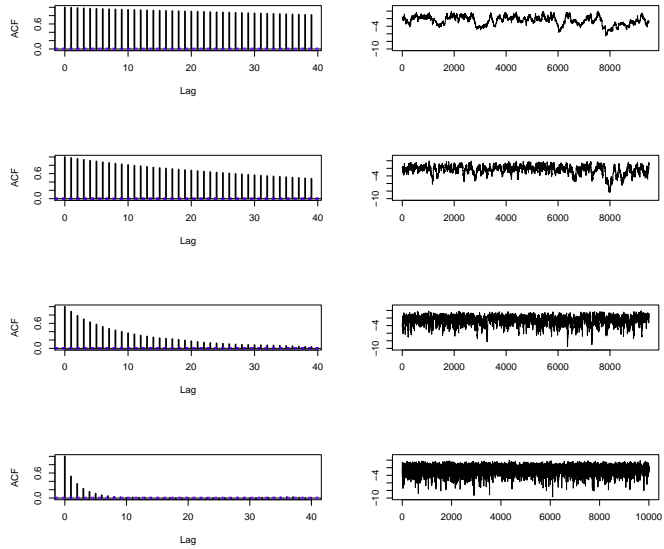}
\caption{Parameter $\beta_1$, model (\ref{toy}). Autocorrelation functions and traces obtained from WinBUGS (first row, based on $\pi(\beta\vert\mathbf{y})$, second 
row, based on orthogonal covariates, third row, based on $\pi(\theta\vert\mathbf{y})$), 
and using our algorithm (last row).}
\label{fig_ejemplo_1_beta1}
\end{figure}

\begin{figure}[h]
\centering
\includegraphics[scale=0.45]{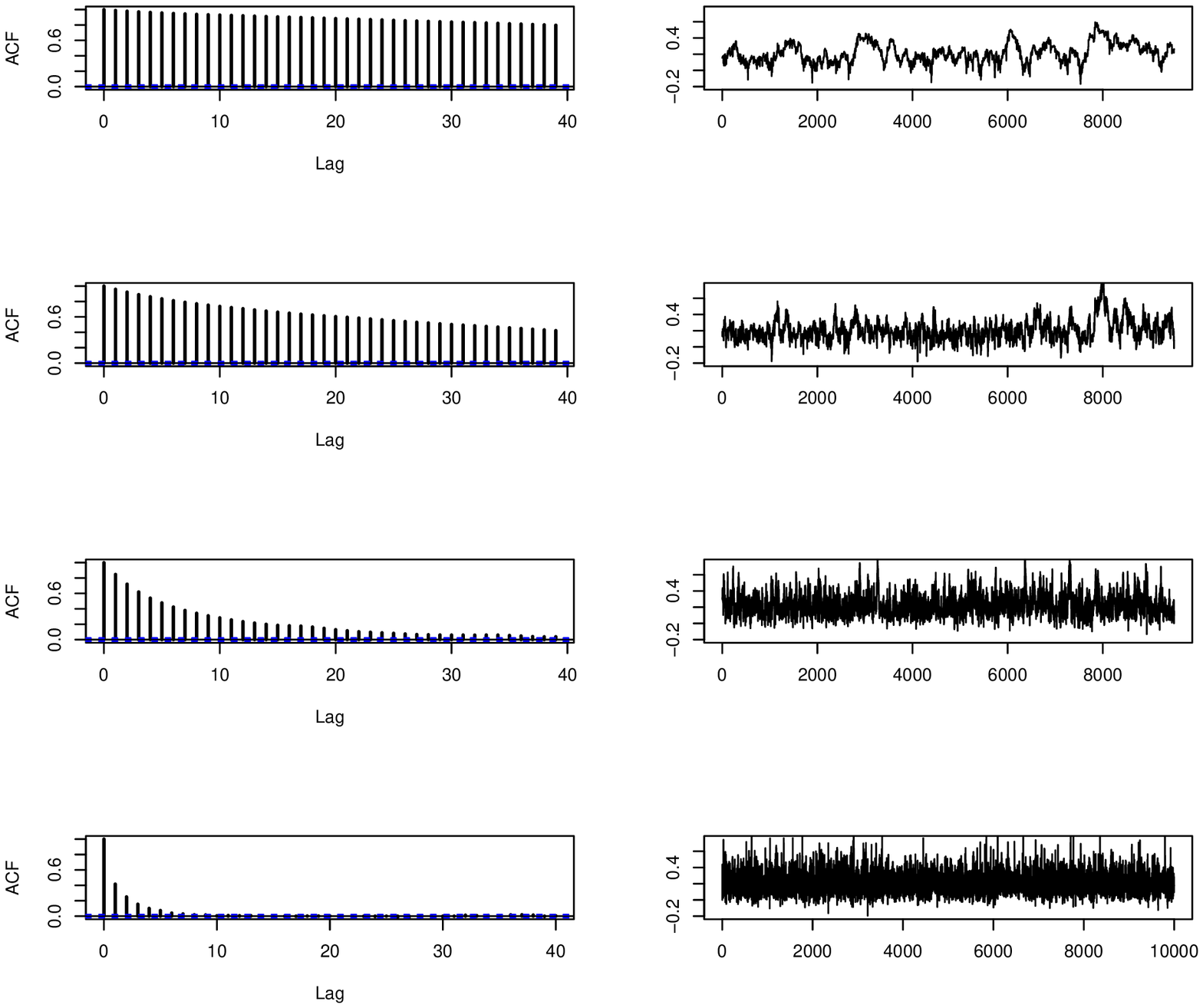}
\caption{Parameter $\beta_2$, model (\ref{toy}). Autocorrelation functions and traces obtained from WinBUGS (first row, based on $\pi(\beta\vert\mathbf{y})$, second 
row, based on orthogonal covariates, and third row, based on $\pi(\theta\vert\mathbf{y})$),
 and using our algorithm (last row).}
\label{fig_ejemplo_1_beta2}
\end{figure}

Figures (\ref{fig_ejemplo_1_beta1}) and (\ref{fig_ejemplo_1_beta2}), first row, show a poor convergence of the chains that may be explained in part by the high posterior correlation (-0.97) between $\beta_1$ and $\beta_2$ and by the constraints. The same results were obtained increasing the adaptive phase to 2000 and using the last 8000 iterations. If we consider orthogonal covariates, that is $x_{i2}=i-5.5$ instead of $x_{i2}=i$, the autocorrelation functions show a moderate improvement (Figures (\ref{fig_ejemplo_1_beta1}) and (\ref{fig_ejemplo_1_beta2}), second row), but a slow convergence again. A better performance would be attained with a reparameterization for which the new parameters were approximately uncorrelated given the data. This reparameterization may be obtained using the estimated covariance matrix of the maximum likelihood estimator of $\beta$. However, very often neither the maximum likelihood estimate nor the estimated covariance matrix can be calculated and this is the first problem we consider. To avoid this drawback we propose a reparameterization based on a Poisson model, see Zou (\citeyear{zou2004}).

\subsection{Reparameterization based on a Poisson model}

Let $\hat{\Sigma}$ be the estimated covariance matrix of the maximum likelihood estimator $\hat{\beta}$ obtained fitting the Poisson model (\ref{Zou_poisson}) and let $L$ be the upper triangular factor of the Choleski decomposition of $\hat{\Sigma}=L^TL$. The likelihood function associated with the log-binomial regression model (\ref{Modelo}) is
\begin{equation}
f(\mathbf{y}|\beta)=\prod_{i=1}^np_i^{y_i}(1-p_i)^{1-y_i},\label{verosi}
\end{equation}
where $p_i=\exp(x_i\beta)$ and $x_i\beta<0$, $i\in\mathbb{N}_n$. The reparameterization we propose is $\theta=L^{-T}\beta$. If $\pi(\beta)$ is the prior distribution then the posterior distribution of $\beta$ is $\pi(\beta|\mathbf{y})\propto\pi(\beta)f(\mathbf{y}|\beta)$ and hence, given the data $\mathbf{y}$, the distribution of $\theta=L^{-T}\beta$ is
\[
\pi(\theta|\mathbf{y})\propto\pi(L^T\theta)\prod_{i=1}^np_i^{y_i}(1-p_i)^{1-y_i},\,\,\,\theta\in\Theta
\]
where now, $p_i=\exp(z_i\theta)$, $z_i=x_iL^T$, $i\in\mathbb{N}_n$ and
\[
\Theta=\{\theta\in\mathbb{R}^k;\,z_i\theta<0\,\,\,\forall i\in\mathbb{N}_n\}.
\]

Using WinBUGS we can simulate a Markov chain with stationary distribution $\pi(\theta|\mathbf{y})$ and therefore this chain can be used to carry out Bayesian inference on $\beta=L^T\theta$. If the posterior distribution of $\beta$ is approximately the multivariate normal distribution $N(\hat{\beta},\hat{\Sigma})$ restricted to $\{\beta\in\mathbb{R}^k;\,x_i\beta<0\,\forall i\in\mathbb{N}_n\}$, then the distribution $\pi(\theta|\mathbf{y})$ is approximately the multivariate normal distribution with mean $\hat{\theta}=L^{-T}\hat{\beta}$ and covariance matrix $L^{-T}\hat{\Sigma}L^{-1}=L^{-T}(L^TL)L^{-1}=\mathbf{I}$, restricted to $\Theta$. Therefore, WinBUGS would get a better convergence if it is used to simulate from $\pi(\theta\vert\mathbf{y})$ instead of directly simulating from the posterior distribution of $\beta$.

For model (\ref{toy}) we have used WinBUGS to simulate from $\pi(\theta\vert\mathbf{y})$.  We have run a chain with 10000 iterations and an adaptive phase of 500 iterations using the method \textit{UpdaterMetnormal}, and the last 9500 iterations have been used to carry out the inferences. After that, we have transformed the simulations using $\beta=L^T\theta$. Figures (\ref{fig_ejemplo_1_beta1}) and (\ref{fig_ejemplo_1_beta2}), third row, show a better convergence of the chains compared with the chains obtained from WinBUGS when the target was $\pi(\beta|\mathbf{y})$. This improvement is due to the reparameterizacion based on the Poisson model.

On the other hand, the methods implemented with WinBUGS to simulate from $\pi(\beta|\mathbf{y})$ or from $\pi(\theta\vert\mathbf{y})$ have the drawback that the instrumental distribution of the Metropolis-Hastings step used to simulate each full conditional distribution has not account for the constraints. Therefore, the constraints must be evaluated \textit{a posteriori} each time a simulation from the instrumental distribution is proposed. This can increase the computational time and the \textit{probability of rejecting} in the Metropolis-Hastings steps. This is the second problem we consider. To overcome it we propose an instrumental distribution designed to efficiently explore the parameter space.

\subsection{Instrumental distribution}
We propose a Metropolis-within-Gibbs algorithm that generates a Markov chain with stationary distribution $\pi(\theta\vert\mathbf{y})$ and therefore it can be used to carry out Bayesian inference on $\beta=L^T\theta$. It is based on an efficient simulation from the full conditional distributions. For $j\in\{1,2,...,k\}$ and $\theta_{\sim j}\in\mathbb{R}^{k-1}$ such that $\pi(\theta_{\sim j}\vert\mathbf{y})=\int\pi(\theta\vert\mathbf{y})d\theta_j>0$, the full conditional distribution is $\pi(\theta_j\vert\mathbf{y},\theta_{\sim j})\propto\pi(\theta|\mathbf{y})$. The set
\[
\Theta_j=\{\theta_j\in\mathbb{R};\pi(\theta_j\vert\mathbf{y},\theta_{\sim j})>0\},
\]
is a key ingredient for our Metropolis-within-Gibbs algorithm. In the following proposition, that is proved in the Appendix, it is established that the set $\Theta_j$ is an interval of the real line.
\begin{proposition}
If $\pi(\theta_{\sim j}\vert\mathbf{y})>0$ then the set $\Theta_j$ is the interval $(a_j,b_j)$ where
\[
a_j=\max_{i\in A_j}\sum_{s\neq j}-z_{is}\theta_s/z_{ij},\,\,\,A_j=\{i\in\mathbb{N}_n;z_{ij}<0\},
\]
and
\[
b_j=\min_{i\in B_j}\sum_{s\neq j}-z_{is}\theta_s/z_{ij},\,\,\,B_j=\{i\in\mathbb{N}_n;z_{ij}>0\},
\]
with the convention that $a_j=-\infty$ if $A_j=\emptyset$ and $b_j=+\infty$ if $B_j=\emptyset$.
\end{proposition}

To get an appropriate instrumental distribution we argue that the multivariate normal distribution $N(\hat{\theta},\mathbf{I})$ restricted to $\Theta$ is an approximation to the distribution $\pi(\theta|\mathbf{y})$ and hence, the distribution $N(\hat{\theta}_j,1)$ restricted to $\Theta_j=(a_j,b_j)$ would be an appropriate instrumental distribution to perform the Metropolis-Hastings step. However, the simulation from a truncated normal distribution can increase the computational time. Instead, we propose the Cauchy distribution with location $\hat{\theta}_j$ and scale $1$ truncated to $\Theta_j$ with density
\begin{equation}
\mathcal{C}(\theta_j^{\prime})\propto\frac{1_{\Theta_j}(\theta_j^{\prime})}
{\pi(1+(\theta_j^{\prime}-\hat{\theta}_j)^2))},\label{cauchy}
\end{equation}
where $1_{\Theta_j}(\theta_j^{\prime})=1$ if $\theta_j^{\prime}\in\Theta_j$ and $0$ otherwise. To simulate $\theta_j^{\prime}$ from this instrumental distribution we simulate $u\sim U(0,1)$ and compute
\[
\theta_j^{\prime}=\hat{\theta}_j-\tan\left((u-1)\arctan(a_j-\hat{\theta}_j)+u\arctan(\hat{\theta}_j-b_j)\right).
\]
The instrumental distribution (\ref{cauchy}) reduces the autocorrelation, as it is shown in the examples. The proposed Metropolis-within-Gibbs algorithm has been implemented in R and it is provided in the Appendix.

Figures (\ref{fig_ejemplo_1_beta1}) and (\ref{fig_ejemplo_1_beta2}), last row, show the results obtained using our MCMC algorithm with 10000 iterations. Our algorithm produces a satisfactory acceptance rate and a quickly decreasing autocorrelation. This improvement, compared with the WinBUGS code used to simulate from $\pi(\theta|\mathbf{y})$ is due to the proposed instrumental distribution (\ref{cauchy}).

\section{Examples}
In this section we present three examples to illustrate our MCMC algorithm.
For each example we have used WinBUGS to simulate from $\pi(\beta|\mathbf{y})$ and from $\pi(\theta\vert\mathbf{y})$ running a chain with an adaptive phase of 500 iterations out of a total of 10000 iterations for the method \textit{UpdaterMetnormal} and after that we have transformed the simulations from $\pi(\theta\vert\mathbf{y})$ using $\beta=L^T\theta$. We have also used our MCMC algorithm with 10000 iterations and the simulations have been transformed using $\beta=L^T\theta$. We have used uniform prior distributions. The efficiency of each algorithm has been measured in terms of the effective sample size and the computational speed.

\subsection{Breast cancer mortality}
We consider the data on the relation between receptor level and stage to 5-year
survival in a cohort of 192 women with breast cancer, see Table (\ref{Greenland}), discussed in \cite{greenland2004}. In this example the percentage of deaths was 28.13\%.

\begin{figure}[h]
\centering
\includegraphics[scale=0.6]{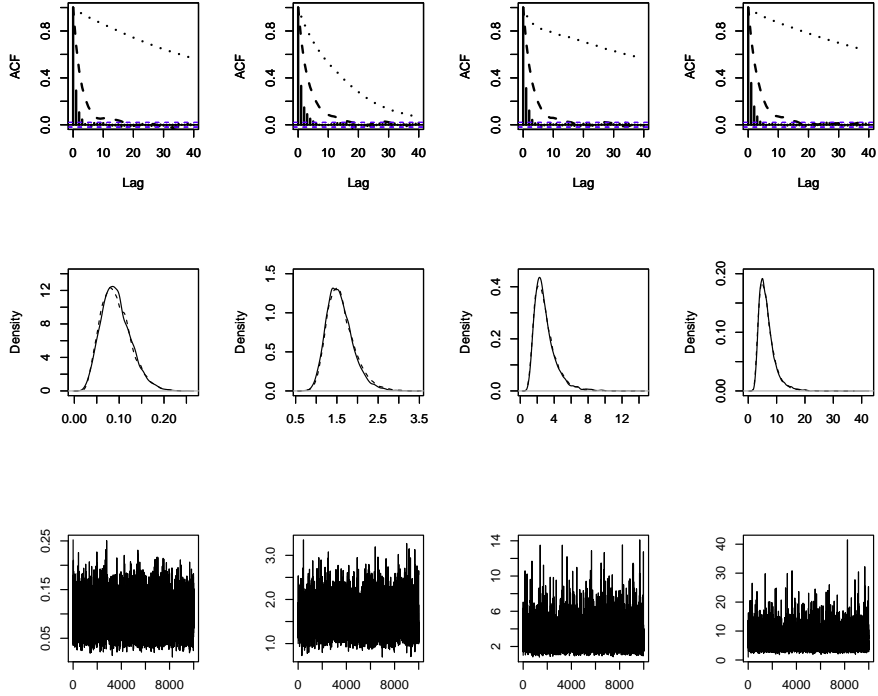}
\caption{Breast cancer mortality example. Parameters $e^{\beta_1}$, $e^{\beta_2}$, $e^{\beta_3}$ and $e^{\beta_4}$ from left to right. First row: autocorrelation functions obtained from WinBUGS with target $\pi(\beta|\mathbf{y})$ (dotted) and $\pi(\theta|\mathbf{y})$ (dashed); and using our algorithm (vertical lines). Second row: posterior densities obtained from WinBUGS with target $\pi(\theta|\mathbf{y})$ (dashed) and obtained with our algorithm (solid). Third row: traces based on our algorithm.}
\label{fig-greeland}
\end{figure}

Figure (\ref{fig-greeland}), first row, shows the autocorrelation functions for the parameters $e^{\beta_1}$, $e^{\beta_2}$, $e^{\beta_3}$ and $e^{\beta_4}$, obtained from WinBUGS with target $\pi(\beta|\mathbf{y})$ (dotted), $\pi(\theta|\mathbf{y})$ (dashed) and using our algorithm (vertical lines). Our algorithm has a satisfactory acceptance rate and a quickly decreasing autocorrelation function. The effective sample sizes are shown in Table (\ref{tabla-ESS-breast}). The results show that our method converges faster than the chains obtained with WinBUGS. Regarding the computational speed, WinBUGS and our MCMC algorithm took seven seconds. Table (\ref{tabla-res-breast}) shows the estimation of the risk ratios obtained with our MCMC algorithm.

\subsection{Low birth weight}
We use the data from a 1986 cohort study conducted at the Baystate Medical Center, Springfield Massachusetts, see \cite{Lemeshow}. The study was designed to identify risk factors associated with an increased risk of low birth weight (weighing less than 2500 grams). Data were collected on 189 pregnant women, 59 of whom had low birth weight infants. We have studied the association between the low birth weight and uterine irritability (ui: yes/no), smoking status during pregnancy (smoke: yes/no), mother's race (race: white, black, other), previous premature labours (ptl$>0$: yes/no), and mother's age (age: $\leq 18$, (18,20], (20,25], (25,30] and $>30$).

\begin{figure}[h]
\begin{center}
\includegraphics[scale=.7]{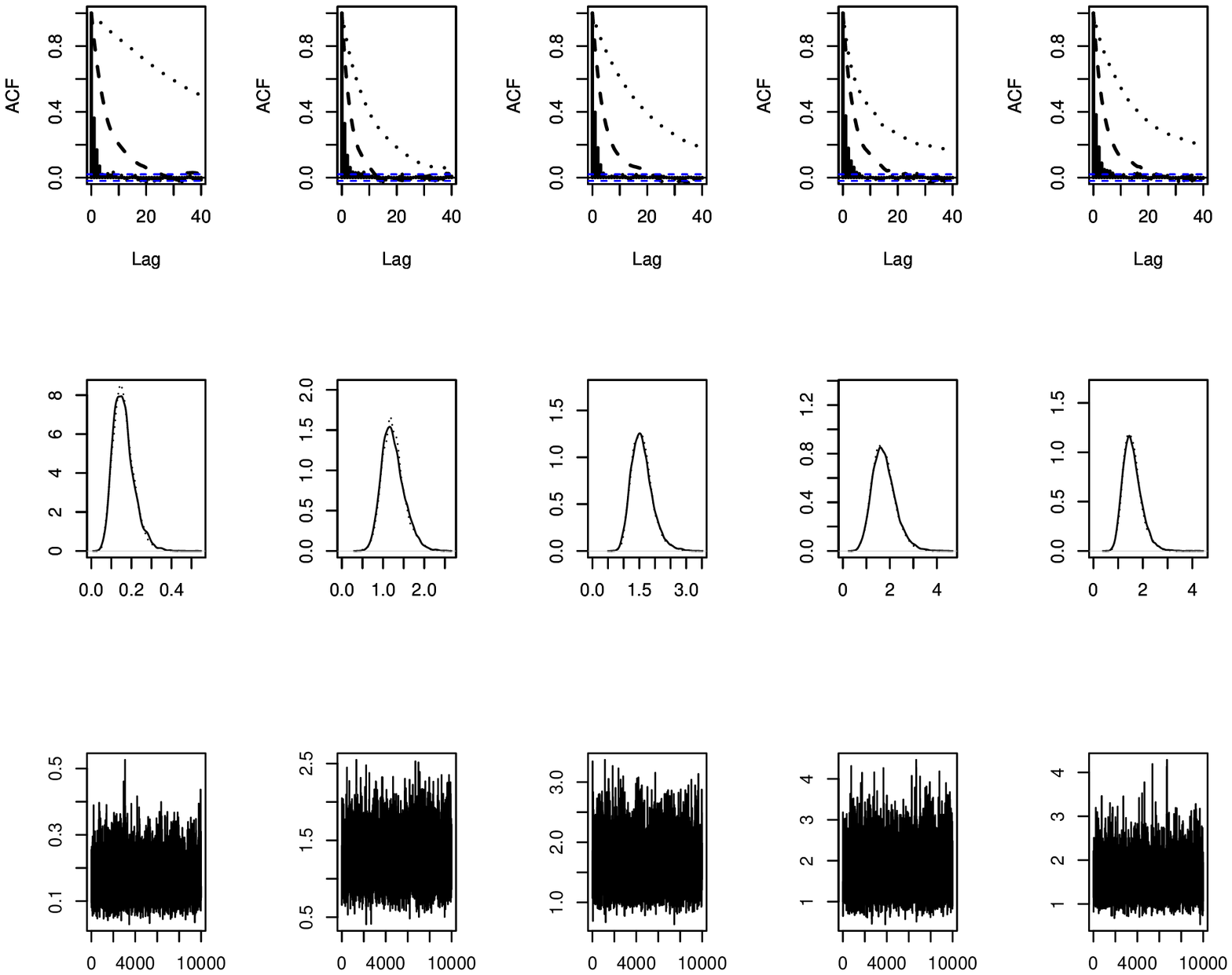}
\caption{Low birth weight example. Parameters $e^{\beta_j}$, $j=1,...,5$ from left to right. First row: autocorrelation functions obtained from WinBUGS with target $\pi(\beta|\mathbf{y})$ (dotted) and $\pi(\theta|\mathbf{y})$ (dashed); and using our algorithm (vertical lines). Second row: posterior densities obtained from WinBUGS with target $\pi(\theta|\mathbf{y})$ (dashed) and obtained with our algorithm (solid). Third row: traces based on our algorithm.}
\label{fig1-pesoNacer}
\end{center}
\end{figure}

\begin{figure}[h]
\begin{center}
\includegraphics[scale=.7]{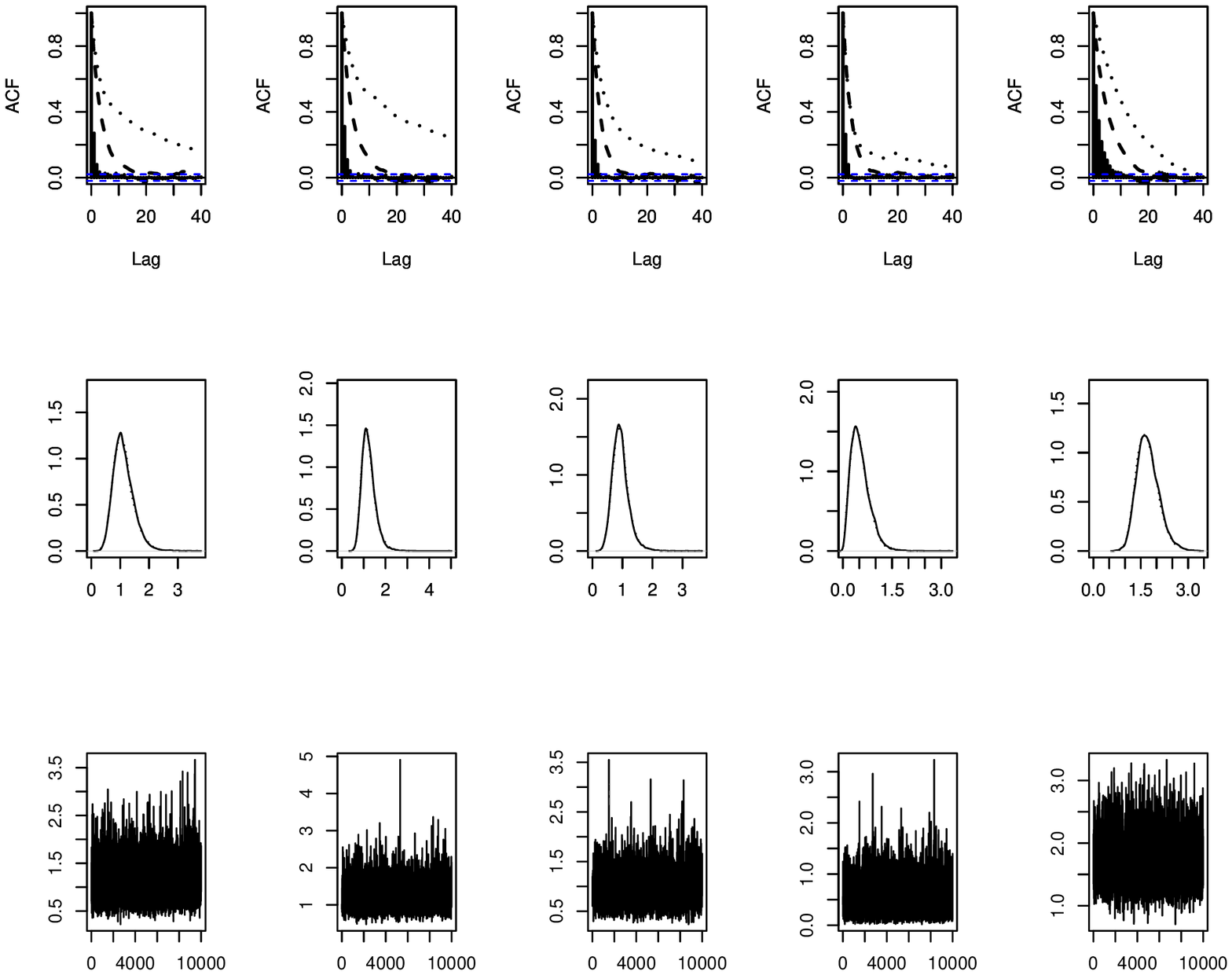}
\caption{Low birth weight example. Parameters $e^{\beta_j}$, $j=6,...,10$ from left to right. First row: autocorrelation functions obtained from WinBUGS with target $\pi(\beta|\mathbf{y})$ (dotted) and $\pi(\theta|\mathbf{y})$ (dashed); and using our algorithm (vertical lines). Second row: posterior densities obtained from WinBUGS with target $\pi(\theta|\mathbf{y})$ (dashed) and obtained with our algorithm (solid). Third row: traces based on our algorithm.}
\label{fig2-pesoNacer}
\end{center}
\end{figure}

Figures (\ref{fig1-pesoNacer}) and (\ref{fig2-pesoNacer}), first row, show the autocorrelation functions for the parameters $e^{\beta_j}$, $j=1,\dots,10$, obtained from WinBUGS with target $\pi(\beta|\mathbf{y})$ (dotted), $\pi(\theta|\mathbf{y})$ (dashed) and using our algorithm (vertical lines). The effective sample sizes are shown in Table (\ref{tabESSpesoNacer}). Again, the results show that our method converges faster than WinBUGS (regarding the computational speed, our MCMC algorithm and WinBUGS took 20 seconds). Table (\ref{tabSUMMARYpesoNacer}) shows the estimates of the risk ratios obtained with our MCMC algorithm. 

\subsection{Simulated example}
We have simulated data form the log-binomial regression model
\[
\log\,p_i=x_i\beta,\,\,\,i\in\mathbb{N}_{1500}=\{1,...,1500\}
\]
where $x_i=(1,x_{i2},...,x_{i9})$ and $x_{ij}$ has been simulated as follows. For $i=1,...,1500$

\begin{itemize}
\item
$
x_{i2}=E_i-1/2,\,
x_{i3}=F_{i1}-1/2,\,
x_{i4}=F_{i2}-1/2,\,
x_{i5}=F_{i3}-1/2,\,
x_{i6}=F_{i4}-1/2,\,
$
where $F_{ij}\sim Ber(1/2)$ for $j=1,2$, $F_{ij}\sim U(0,1)$ for $j=3,4$ and the distribution of $E_i$ is the Bernoulli distribution
\[
Ber(\exp(\alpha_1+\alpha_2(F_{i1}-1/2)+\alpha_3(F_{i2}-1/2)+\alpha_4(F_{i3}-1/2)+\alpha_5(F_{i4}-1/2))
\]

\item We have simulated $(w_{i1},w_{i2})$ from a multivariate normal distribution with mean $(0,0)$, $Var(w_{i1})=Var(w_{i2})=1$ and $Cov(w_{i1},w_{i2})=0.5$. Then we have calculated
$\tilde{w}_{ij}=w_{ij}-\min_iw_{ij}$ and $V_{ij}=\tilde{w}_{ij}/\max_i\tilde{w}_{ij}$, $j=1,2$, and $V_{i3}\sim U(0,1)$. Finally,
\[
x_{i7}=V_{i1}-1/2,\,
x_{i8}=V_{i2}-1/2,\,
x_{i9}=V_{i3}-1/2.
\]
\end{itemize}

The value of the parameters $(e^{\beta_1},...,e^{\beta_9})$ and $(e^{\alpha_1},...,e^{\alpha_5})$ used to simulate the data $\mathbf{y}$ were
\[
(0.379, 1.400, 1.200, 1.300, 1.100, 1.250, 1.500, 1.400, 1.100).
\]
and
\[
(0.512, 1.400, 1.200, 1.600, 1.400),
\]
respectively. Thus, $E$ may represent an exposure, $F_1$, $F_2$, $F_3$ and $F_4$ confounders and $V_1$, $V_2$ and $V_3$ predictors. With this value of $\beta$ we have computed $p_i=\exp(x_i\beta)$ and we have simulated the outcome $y_i\sim Ber(p_i)$, $i\in\mathbb{N}_{1500}$, obtaining $\overline{y}=\sum_i y_i/n=0.39$.

\begin{figure}[h]
\begin{center}
\includegraphics[scale=.7]{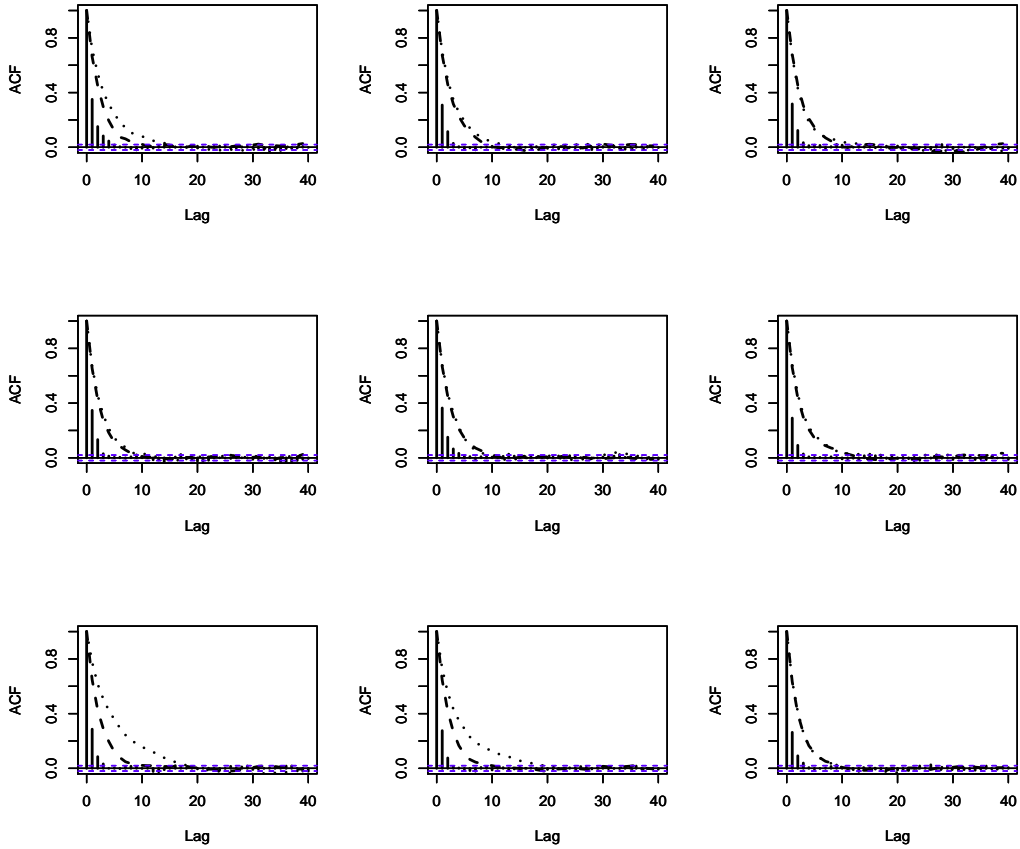}
\caption{MCMC output for the simulated data. Parameters $e^{\beta_j}$, $j=1,...,9$. Autocorrelation functions obtained from WinBUGS with target $\pi(\beta|\mathbf{y})$ (dotted) and $\pi(\theta|\mathbf{y})$ (dashed); and using our algorithm (vertical lines).}
\label{fig1}
\end{center}
\end{figure}

\begin{figure}[h]
\begin{center}
\includegraphics[scale=.45]{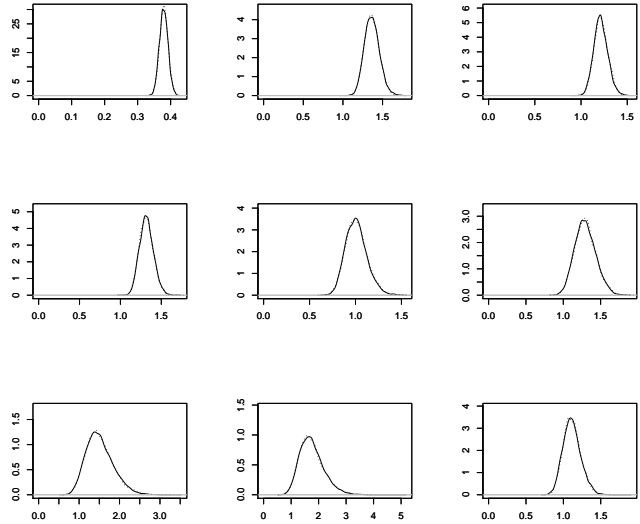}
\caption{MCMC output for the simulated data. Parameters $e^{\beta_j}$, $j=1,...,9$. Posterior densities obtained from WinBUGS with target $\pi(\theta|\mathbf{y})$ (dotted) and obtained with our algorithm (solid).}
\label{fig2}
\end{center}
\end{figure}

\begin{figure}[h]
\begin{center}
\includegraphics[scale=.45]{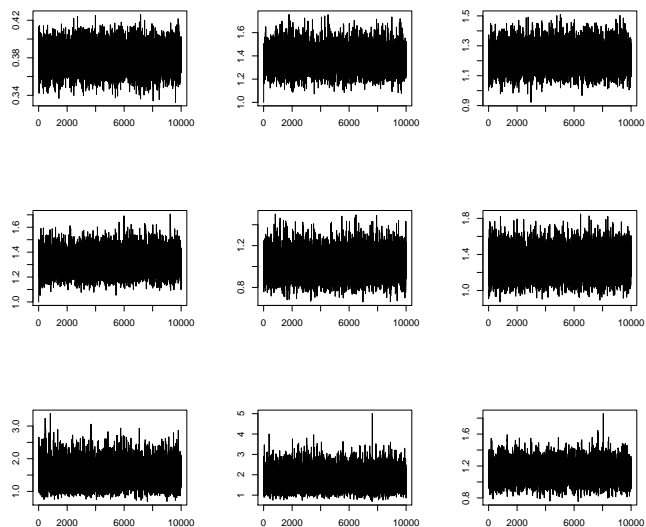}
\caption{MCMC output for the simulated data. Parameters $e^{\beta_j}$, $j=1,...,9$. Traces based on our algorithm.}
\label{fig3}
\end{center}
\end{figure}

Figure (\ref{fig1}) shows the autocorrelation functions obtained from WinBUGS with target $\pi(\beta|\mathbf{y})$ (dotted), $\pi(\theta|\mathbf{y})$ (dashed) and using our algorithm (vertical lines). For some parameters the autocorrelation functions obtained from WinBUGS are virtually identical, and for other parameters, WinBUGS with target $\pi(\theta|\mathbf{y})$ converges faster than WinBUGS with target $\pi(\beta|\mathbf{y})$. For all the parameters our proposed MCMC method produced a satisfactory acceptance rate (see Figures \ref{fig2} and \ref{fig3}) and it was superior to the method implemented with WinBUGS. Regarding the computational speed, our MCMC algorithm took between 86 and 87 seconds while WinBUGS took between 163 and 193 seconds. The effective sample sizes are shown in Table (\ref{tabla_2}). Table (\ref{tabSUMMARYEjemploSimulado}) shows the estimation of the risk ratios obtained with our MCMC algorithm. The posterior mean and the 95\% CI for $e^{\beta_1}$ were 0.379 and (0.354, 0.405), respectively.

\section{Conclusions}
Despite recent efforts made by several authors, logistic regression is still used frequently in cohort studies and clinical trials with common outcome and equal follow-up times, even if one wants to communicate a risk ratio. It is well known that the more frequent the outcome is the more the odds ratio overestimates the risk ratio when it is greater than 1 (or underestimates it if it is less than 1).

If one wants to estimate an adjusted risk ratio, the log-binomial model is preferable to the logistic one but the constrained parameter space makes  difficult to find the maximum likelihood estimate. Bayesian methods implemented with WinBUGS can work with a constrained parameter space in a natural way. Moreover, \cite{chucole} have shown that Bayesian methods produce smaller mean squared errors than likelihood based methods. However, WinBUGS can lead to a high Monte Carlo error. 

To avoid this drawback, we have proposed an efficient MCMC algorithm to estimate risk ratios from a Bayesian point of view using log-binomial regression models. Our method is based on two strategies: first, a reparameterization based on a Poisson model, and second, an appropriate Cauchy instrumental distribution. It converges to the posterior distribution faster than the methods implemented with WinBUGS. Regarding the computational speed, our MCMC algorithm is similar to WinBUGS for moderate sample sizes and faster for large sample sizes. Furtheremore, the possibility of easily carrying out the estimations using our R functions is an important added value.
\newline

\section*{Acknowledgment}
This research was supported by the S\'eneca Foundation Programme
for the Generation of Excellence Scientific Knowledge under
Project 15220/PI/10.
\par

\newpage

\section*{Appendix}

\subsubsection*{WinBUGS code for model (\ref{toy}) proposed by \cite{chucole} to simulate from $\pi(\beta|\mathbf{y})$}
\begin{verbatim}
model{
for(i in 1:N){
 p[i]<-exp(beta1+i*beta2)
 y[i]~dbern(p[i])
}
beta1~flat()
beta2~flat()
for(i in 1:N){
 ones[i]<-1
 ones[i]~dbern(q[i])
 q[i]<-step(1-p[i])}
}
\end{verbatim}

\subsubsection*{Proof of proposition 1}
Because of $\pi(\theta_{\sim j}\vert\mathbf{y})>0$, there exist $\theta_j^*\in\mathbb{R}$ such that $\pi(\theta_j^*,\theta_{\sim j}\vert\mathbf{y})>0$ and hence $\theta_j^*\in\Theta_j$. Using that $\pi(\theta_j^*,\theta_{\sim j}\vert\mathbf{y})>0$ it follows that $z_{ij}\theta^*_j+\sum_{s\neq j} z_{is}\theta_s<0$ for $i\in\mathbb{N}_n$ and then
\[
\sum_{s\neq j} z_{is}\theta_s<0,\,\forall i\in\mathbb{N}_n\,\,\, \mathrm{such}\,\mathrm{that}\,z_{ij}=0.
\]

Let $\theta_j$ be a real number. Then $\theta_j\in\Theta_j$ if and only if
\[
z_{ij}\theta_j+\sum_{s\neq j} z_{is}\theta_s<0,\,\forall i\in\mathbb{N}_n,
\]
that is, if and only if
\[
\theta_j>\sum_{s\neq j}-z_{is}\theta_s/z_{ij},\,\forall i\in A_j,
\]
\[
\theta_j<\sum_{s\neq j}-z_{is}\theta_s/z_{ij},\,\forall i\in B_j
\]
and
\[
\sum_{s\neq j} z_{is}\theta_s<0,\,\forall i\in\mathbb{N}_n\,\,\, \mathrm{such}\,\mathrm{that}\,z_{ij}=0.
\]
It follows that $\Theta_j= (a_j,b_j)$.

\subsubsection*{R functions}

\begin{verbatim}

gibbsLogBinomial=function(j){

ztheta=Z[,-j]%*%matrix(theta[-j],ncol=1)

A=Aind[[j]];B=Bind[[j]]

suma1=sum(Z[,j]<0);a=-Inf
if(suma1!=0){a=max(-ztheta[A]/Z[A,j])}

suma2=sum(Z[,j]>0);b=Inf
if(suma2!=0){b=min(-ztheta[B]/Z[B,j])}

u=runif(1,0,1)
location=theta.hat[j]
thetaj.star=location-tan((u-1)*atan(a-location)+u*atan(location-b))

theta.new=theta;theta.new[j]=thetaj.star

p.new=exp(Z[,j]*(thetaj.star-theta[j]))*p

logvalue.new=sum(log(p.new[y==1]))+sum(log(1-p.new[y==0]))
priortheta.new=prior(theta.new)

rho=exp(logvalue.new-logvalue)
rho=rho*priortheta.new/priortheta
rho=rho*(1+(thetaj.star-location)^2)/(1+(theta[j]-location)^2)
rho=min(1,rho)

logvalue<<-logvalue
theta<<-theta
p<<-p
priortheta<<-priortheta

u=runif(1,0,1)
if(u<rho){theta<<-theta.new;logvalue<<-logvalue.new;p<<-p.new;
priortheta<<-priortheta.new}

}

prior=function(theta){return(1)}

inicial.beta=function(){
coef=summary(glm(y ~ 1,family=binomial))$coeff
mu=coef[1,1];serror=coef[1,2]
musim=rnorm(1,mu,serror)
beta1=log(exp(musim)/(1+exp(musim)))
return(c(beta1,rep(0,k-1)))}

initialize=function(){

#Reparameterization

X<<-model.ini$x;n<<-nrow(X);beta=model.ini$coeff
Sigma<<-summary(model.ini)$cov.unscaled
L<<-chol(Sigma)
Z<<-X%*%t(L)
model.ini.0<<-glm(y ~ Z-1,family=poisson,x=TRUE)
theta.hat<<-solve(t(L))%*%beta;k<<-ncol(Z)

#Sets in proposition 1

Aind<<-{}
for(j in 1:k){ 
Aind[[j]]<<-(1:n)[Z[,j]<0]}
Bind<<-{}
for(j in 1:k){ 
Bind[[j]]<<-(1:n)[Z[,j]>0]}

#Initial point. The following lines are always the same, although 
#the user can change punto to an other inital point

punto<<-solve(t(L))%*%inicial.beta()
theta<<-punto
p<<-exp(Z%*%theta)
logvalue<<-sum(log(p[y==1]))+sum(log(1-p[y==0]))
priortheta<<-prior(theta)
}
\end{verbatim}

\subsubsection*{Using the R function \textsf{gibbsLogBinomial} with the breast cancer mortality example}
\begin{verbatim}
#The data

datos<-rbind(cbind(rep(1,12),rep(1,12),c(rep(1,2),rep(0,10))),
cbind(rep(1,55),rep(2,55),c(rep(1,5),rep(0,50))),
cbind(rep(2,22),rep(1,22),c(rep(1,9),rep(0,13))),
cbind(rep(2,74),rep(2,74),c(rep(1,17),rep(0,57))),
cbind(rep(3,14),rep(1,14),c(rep(1,12),rep(0,2))),
cbind(rep(3,15),rep(2,15),c(rep(1,9),rep(0,6))))

datos<-data.frame(datos)
names(datos)<-c("Stage","Receptor_Level","Dead")

#Recoding Receptor_level

datos$Receptor_Level=as.integer(datos$Receptor_Level==1)

#Outcome

y=datos$Dead

##############################################################
##############################################################

################ Runing the MCMC algorithm ###################

#Poisson model. The following line depends on covariates


model.ini=glm(y~factor(Receptor_Level)+factor(Stage),
family=poisson,data=datos,x=TRUE)

#The following lines compute the need input for 
#the algorithm and fix the lengtht of the chain to 10000

initialize()
longChain=10000
theta.sim=matrix(rep(NA,longChain*k),ncol=k)

#Finally the chain is simulated as follows

for(h in 1:longChain){
 theta.sim[h,]=theta
   for(j in 1:k){
     gibbsLogBinomial(j)
   }
}
beta.sim=theta.sim%*%L

#The object beta.sim containts the simulations
#Posterior estimation of exp(beta) using the coda package

library(coda)

RR=mcmc(exp(beta.sim))
summary(RR)
autocorr.plot(RR)
effectiveSize(RR)
plot(RR)
\end{verbatim}

\newpage

\begin{table}[h]
\centering \caption{Data relating receptor level (low (1) and high(2)) and stage to
5-year breast cancer mortality.}
\begin{tabular}{cccc}
\\
 \hline
  Stage & Receptor level & Deaths & Total \\
  \hline
1 & 1 & 2 & 12\\

1 & 2 & 5 & 55\\

2 & 1 & 9 & 22\\

2 & 2 & 17 & 74\\

3 & 1 & 12 & 14\\

3 & 2 & 9 & 15\\
\hline
\end{tabular}
\label{Greenland}
\end{table}

\begin{table}[h]
\centering
\caption{Breast cancer mortality example. Effective sample sizes obtained with our algorithm (first row), WinBUGS with target $\pi(\theta|\mathbf{y})$ (second row) and WinBUGS with target $\pi(\beta|\mathbf{y})$ (third row).}
\begin{tabular}{cccc}\\
\hline\\[-1.5ex]
$e^{\beta_1}$ & $e^{\beta_2}$ & $e^{\beta_3}$ & $e^{\beta_4}$\\
\hline
5636.9    &    4464.8    &    5450.6    &    4685.2\\
1829.3   &     1360.8    &    1600.2   &     1433.4\\
70.7    &     324.8     &     61.2     &     59.4\\
\hline
\end{tabular}
\label{tabla-ESS-breast}
\end{table}

\begin{table}[h]
\centering
\caption{Bayesian estimation of the risk ratios obtained using our MCMC algorithm for the breast cancer mortality data: posterior mean, $E(RR|\mathbf{y})$, and 95\% credible interval (95\% CI).}
\begin{tabular}{lccc}
\\
\hline
 & $E(RR|\mathbf{y})$ &  95\% CI\\
\hline
receptor low  & 1.576 & (1.041, 2.364)\\
stage 2 & 2.939 & (1.256,  6.404)\\
stage 3 & 6.626 & (2.871, 14.258)\\
\hline
\end{tabular}
\label{tabla-res-breast}
\end{table}

\begin{table}[h]
\centering
\caption{Bayesian estimation of the risk ratios obtained using our MCMC algorithm for the low birth weight data: posterior mean, $E(RR|\mathbf{y})$, and 95\% credible interval (95\% CI).}
\begin{tabular}{lccc}
\\
\hline
 & $E(RR|\mathbf{y})$ &  95\% CI\\
\hline
ui yes & 1.242 & (0.780, 1.863)\\
smoke yes & 1.586 & (1.022, 2.377)\\
race black & 1.757 & (0.926, 2.893)\\
race other & 1.573 & (0.969, 2.439)\\
age (18,20] & 1.120 & (0.554, 1.921)\\
age (20,25] & 1.226 & (0.739, 1.944)\\
age (25,30] & 0.934 & (0.485, 1.574)\\
age $> 30$ & 0.532 &  (0.115, 1.199)\\ 
ptl$>0$ yes & 1.727 & (1.133, 2.514)\\
\hline
\end{tabular}
\label{tabSUMMARYpesoNacer}
\end{table}

\begin{table}[h]
\centering
\caption{Low birth weight example. Effective sample sizes obtained with our MCMC algorithm (rows a and d), WinBUGS with target $\pi(\theta|\mathbf{y})$ (rows b and e) and WinBUGS with target $\pi(\beta\vert\mathbf{y})$ (rows c and f).}
\begin{tabular}{cccccc}\\
\hline\\[-1.5ex]
 & $e^{\beta_1}$ & $e^{\beta_2}$ & $e^{\beta_3}$ & $e^{\beta_4}$ & $e^{\beta_5}$\\
\hline
a & 4239.9 & 4756.9 & 4084.8 & 3965.8 & 4074.3\\
b & 852 & 1426.3 & 1067.7 & 956.9 & 1161.9\\
c & 85.1 & 449.9 & 235.8 & 259.5 & 276.9\\

\hline\\[-1.5ex]
 & $e^{\beta_6}$ & $e^{\beta_7}$ & $e^{\beta_8}$ & $e^{\beta_9}$ & $e^{\beta_{10}}$\\
\hline
d & 5753.7 & 5786.8 & 5937.3 & 5837.2 & 2414.2\\
e & 1283.1 & 1163.8 & 1563 & 1560.7 & 845.6\\
f & 318.4 & 243.5 & 486.5 & 693.5 & 367.1\\
\hline
\end{tabular}
\label{tabESSpesoNacer}
\end{table}

\begin{table}[h]

\centering

\caption{Bayesian estimation of the risk ratios obtained using our MCMC algorithm for the simulated data: posterior mean, $E(RR|\mathbf{y})$, and 95\% credible interval (95\% CI).}

\begin{tabular}{lcc}

\\

\hline
 & $E(RR|\mathbf{y})$ &  95\% CI\\

\hline
$x_2$ & 1.367 & (1.195, 1.565)\\
$x_3$ & 1.214 & (1.073, 1.372)\\
$x_4$ & 1.320 & (1.164, 1.497)\\
$x_5$ & 1.008 & (0.807, 1.259)\\
$x_6$ & 1.295 & (1.044, 1.588)\\
$x_7$ & 1.511 & (0.955, 2.244)\\
$x_8$ & 1.755 & (1.046, 2.765)\\
$x_9$ & 1.115 & (0.902, 1.371)\\
\hline

\end{tabular}

\label{tabSUMMARYEjemploSimulado}

\end{table}

\begin{table}[h]
\centering
\caption{Simulated example. Effective sample sizes obtained with our MCMC algorithm (first row), WinBUGS with target $\pi(\theta|\mathbf{y})$ (second row) and WinBUGS with target $\pi(\beta|\mathbf{y})$ (thrid row).}
\begin{tabular}{ccccccccc}\\
\hline\\[-1.5ex]
$e^{\beta_1}$ & $e^{\beta_2}$ & $e^{\beta_3}$ & $e^{\beta_4}$ & $e^{\beta_5}$ & $e^{\beta_6}$ & $e^{\beta_7}$ & $e^{\beta_8}$ & $e^{\beta_9}$\\ 
\hline

4299.5 & 5058.5 & 4930.9 & 4979 & 4466.4 & 5495.6 & 5546.3 & 5677.3 & 5557\\
1911 & 1854.2 & 1822.1 & 2002 & 1899.7 & 1951.8 & 1914.5 & 2052 & 2136\\
1405.4 & 1684.7 & 1934.6 & 1818 & 1841.3 & 1799.6 & 975 & 1141.7 & 2050\\

\hline
\end{tabular}
\label{tabla_2}
\end{table}


\begin{thebibliography}{}

\bibitem[Chu and Cole, 2010]{chucole}
Chu, H., Cole, SR. (2010). Estimation of Risk Ratios in Cohort Studies With Common Outcomes: a Bayesian approach. Epidemiology, 21: 855-862.


\bibitem[Deddens \textit{et al.}, 2003]{pedersen2003} Deddens, J.A., Petersen, M.R., Lei, X. (2003). Estimation of prevalence ratios when PROC GENMOD does not converge. Proceedings of the 28th Annual SAS Users Group International Conference, Seattle, Washington.

\bibitem[Deddens and Petersen, 2008]{deddens} Deddens, J.A., Petersen, M.R. (2008) Approaches for estimating prevalence ratios.
Occupational and Environmental Medicine, 65:501-506.

\bibitem[Greenland, 2004]{greenland2004}
Greenland, S. (2004). Model-based Estimation of Relative Risks 
and Other Epidemiologic Measures in Studies of Common Outcomes and in Case-Control Studies. American Journal of Epidemiology, 160: 301--305

\bibitem[Hosmer and Lemeshow, 2000]{Lemeshow}
Hosmer, D. W. and Lemeshow, S. (2000). Applied Logistic Regression, 2nd edition. New York: John Wiley and Sons. 

\bibitem[McNutt \textit{et al.}, 2003]{mcnutt}
McNutt, LA., Wu, C., Xue, X. and Hafner, JP. (2003). Estimating the Relative Risk in Cohort Studies and Clinical Trials of Common Outcomes. American Journal of Epidemiology, 157: 940--943.


\bibitem[Petersen and Deddens, 2006]{pedersen} Petersen, M.R., Deddens, J.A. (2006). RE: "Easy SAS calculations for risk or
prevalence ratios and differences". American Journal of Epidemiology,163:1158--
1159.

\bibitem[Spiegelman and Hertzmark, 2005]{Spiegelman2005} Spiegelman, D. and Hertzmark, E. (2005). Easy SAS calculations for risk or prevalence
ratios and differences. American Journal of Epidemiology, 162:199--200.


\bibitem[Zou, 2004]{zou2004} Zou, GY. (2004). A modified Poisson regression approach to prospective studies with binary data. American Journal of Epidemiology, 159:702--706.

\end{thebibliography}
\end{document}